\documentclass[useAMS,usenatbib]{mn2e} 
\def\e{$\pm$} 

\def\kms{km~s$^{-1}$} 
\def\vsi{$v\: \sin i$} 
\def\rsun{R_{\odot}}
\def\lsun{L_{\odot}}
\def\msun{M_{\odot}}

\title[Rotational velocities of the giants in symbiotic stars: I. D'--type ]
      {Rotational velocities of the giants in symbiotic stars: \\ I. D'--type symbiotics 
      \thanks{based on observations obtained in ESO programs 073.D-0724A
              and 074.D-0114}} 
\author[Zamanov, Bode, Melo, et al. ] {
R. K. Zamanov$^{1,2}$  \thanks{e-mail: rkz@astro.bas.bg; mfb@astro.livjm.ac.uk; andreja.gomboc@fmf.uni-lj.si}, 
M. F. Bode$^{1}$, 
C. H. F. Melo$^{3,4}$,
J. Porter$^{1}$,\\
\\
\LARGE
{\rm A. Gomboc$^{1,5}$,  R. Konstantinova-Antova$^2$ } \\
\\
$^1$ Astrophysics Research Institute, Liverpool John Moores University, 
       Twelve Quays House, Birkenhead, CH41 1LD, UK\\ 
$^2$ Institute of Astronomy, Bulgarian Academy of Sciences, 
       72 Tsarighradsko Shousse Blvd., 1784 Sofia, Bulgaria \\ 
$^3$ European Southern Observatory, Casilla 19001, Santiago 19, Chile \\
$^4$ Departamento de Astronom\'{\i}a, Universidad de Chile, Casilla 36-D, Santiago, Chile  \\
$^5$ Department of Physics, University of Ljubljana, 
       Jadranska 19, 6100 Ljubljana, Slovenia
} 
\begin{document} 
\date{Accepted . Received 2005 July 30; in original form 2005 July 30} 
 \pagerange{\pageref{firstpage}--\pageref{lastpage}} \pubyear{2005} 
 
\maketitle 
 
\label{firstpage} 
 
\begin{abstract} 
We have measured the rotational velocities (\vsi) of the mass donors in a number of 
D'--type symbiotic stars, using the cross-correlation function method. 
Four from five D' symbiotic stars with known  \vsi,  
appeared to be very fast rotators compared 
with the catalogues of \vsi\ for the corresponding spectral types.
At least three of these stars rotate at a substantial fraction ($\ga0.5$) of
the critical velocity.
This means that at least in D'--type SS the cool components rotate
faster than isolated giants.
If these binary stars are synchronized, their orbital periods 
should be relatively short (4-60 days).
We also briefly discuss the possible origin of the rapid rotation
and its connection with mass loss and dust formation. 

\end{abstract} 
 
\begin{keywords} 
stars: binaries: symbiotic -- stars: rotation -- stars: late type
\end{keywords} 
 
\section{Introduction} 
The Symbiotic stars (SSs -- thought to comprise  a white dwarf                   
(WD) accreting from a  cool giant or Mira) represent the 
extremum of the interacting binary star classification. They                  
offer a unique and exciting laboratory in which to study                   
such important processes  as (i) mass loss from cool giants
and the formation of Planetary Nebulae; (ii) accretion onto                   
compact objects, (iii) photoionisation  and radiative transfer                   
in gaseous nebulae, and (iv) nonrelativistic jets and bipolar
outflows (i.e. Kenyon 1986; Corradi et al. 2003).
                   
Soker (2002) has shown theoretically that the cool companions in symbiotic systems 
are likely to rotate much faster than isolated cool giants or those in wide binary systems. 	   
However, there are no systematic investigations of \vsi\ of the mass donors in SS.
		   
On the basis of their IR properties, SS have been classified into stellar continuum (S)
and dusty (D or D') types. The D--type systems contain Mira variables as mass donors. 
The D'--type  are 
characterized by an earlier spectral type (F-K) of the cool component.
and lower  dust temperatures. Among 188 objects in the latest catalogue of 
symbiotic stars (Belczy{\' n}ski et al. 2000), 
there are only seven that are classified as D'--type:
Wray15-157, AS 201, V417 Cen, HD~330036, AS 269, StH$\alpha$~190
and  Hen 3-1591 (Hen 3-1591 can be S or D').
Three of these have been studied using model atmospheres 
and all display rapid rotation and s-process elemental over-abundances 
(see Pereira et al. 2005).

Our aims here are:
{\bf (1)} to measure the projected rotational velocities (\vsi) 
and  the rotational periods
(P$_{rot}$) of the  giants in  D' symbiotic stars, 
using a cross correlation function (CCF) approach; 
{\bf (2)} to test the theoretical predictions that 
the mass donors in SSs are
faster rotators than  the isolated giants or those 
in wide binary systems; 
{\bf (3)} to provide pointers to the determination of binary periods (assuming co-rotation). This is the first of a series of papers exploring the rotation velocities of the mass donating (cool) components of SSs. 
		                                    
\section{ Observations}
                   
The observations were performed  with FEROS at the 2.2m 
telescope (ESO, La Silla).   
FEROS is a fibre-fed Echelle spectrograph, providing a high resolution of 
$\lambda/\Delta \lambda=$48000, 
a wide wavelength coverage from about 4000~\AA\  to 8000~\AA\  in one exposure 
and a high overall efficiency (Kaufer et al. 1999). 
The 39 orders of the Echelle spectrum are registered on a 2k$\times$4k EEV CCD. 
All spectra are reduced using the dedicated FEROS data reduction software 
implemented in the ESO-MIDAS system.

\begin{table*}  
\caption{Journal of observations and projected rotational velocities of D' type SSs
(note that value for AS~201 is from the literature).  
The spectral types of the giants are from different papers (see Section \ref{indiv}),
(B-V)$_0$ is the intrinsic colour for the corresponding spectral type.
\vsi\ is the rotational velocity  measured in this paper.
The last column gives other measurements of \vsi\ (if available).}  
  \begin{tabular}{@{}llllccccc@{}}  
\hline
object      & date-obs    &    MJD    & Exp. time          &  Cool Star   &(B-V)$_0$& \vsi\     & other	&  \\ 
	    &             &           &                    & Spectrum&         & [\kms ]   & [\kms ]	&  \\
\hline							   		     
HD 330036       & 2004-05-22 & 53147.212 &  2$\times$10 min&  F8III  &  0.90&  107.0\e10   & 100\e10$^a$ & \\ 	
Hen 3-1591      & 2004-07-19 & 53205.222 &  2$\times$20 min&  K1III  &  1.09&  23.71\e 1.5 &	         & \\
StH$\alpha$ 190 & 2004-06-03 & 53159.397 &  2$\times$10 min&G4III/IV &  0.88&  105.0\e10   & 100\e10$^b$ & \\
V417 Cen        & 2004-04-14 & 53109.280 &  2$\times$10 min& G9Ib-II &  0.98&  75.0\e 7.5  &	         & \\  
AS~201          & ........   & ......    &  ........       & F9III   &  0.58&  ......      &25\e5$^a$    & \\ 	 
 \hline 						 
 \label{tab-jou} 
% \hline       

$^a$Pereira et al.(2005);  $^b$Smith  et al.(2001); 				   				      
  \end{tabular}									   				      
  \end{table*}	  

\section{\vsi\ measurement technique}
%  \subsection{Projected rotational velocity - \vsi}
  \label{prv}

Radial velocities and projected rotational velocities have been derived
by cross-correlating the observed spectra with a K0-type numerical mask yielding a
cross-correlation function (CCF) whose centre gives the radial velocity and whose width is related to the
broadening mechanisms affecting the whole spectra such as stellar rotation and turbulence.
Details of the cross-correlation procedure are given in Melo et al. (2001). 

In order to use the width ($\sigma$) of the CCF as an estimate of \vsi\ one needs to subtract
the amount of broadening contributing to $\sigma$ unrelated to stellar rotation (e.g., 
convection, instrumental profile, etc.),
i.e., $\sigma_0$. 
Melo et al. (2001) calibrated $\sigma_0$ as a function of 
$(B-V)$ for FEROS spectra of stars with $0.6 < (B-V) < 1.2$.
The \vsi\ measured from our CCFs for a set of standard stars within
this $B-V$ range are in good agreement with the literature values.
Therefore, for all 4 giants in Table~\ref{tab-jou}, 
%with $(B-V)<1.2$ 
the Melo et al. (2001) calibration 
has been adopted.

For \vsi\ greater than about 30 \kms\, the shape of the CCF becomes gradually closer 
to the Gray rotational profile (Gray 1976). Therefore in order to correctly fit the 
CCF, a different approach is needed as described in Melo (2003). 
The CCF is then fitted by a family of functions $CCF_{V\sin i}=C-D[g_0\otimes G(V\sin i)]$ which
is the result of the  convolution of the CCF of a
non-rotating star $g_0$, which can be fairly approximated by a gaussian, and
the Gray (Gray 1976) rotational profile computed for several rotational 
velocities $G(V\sin i)$. For each function
$CCF_{V\sin i}$ we found the radial
velocity $V_r$, the depth $D$ and the continuum $C$ for minimizing the quantity $\chi^2_{V\sin i}$
which is the traditional $\chi^2$ function with $\sigma_i=1$, where
$\sigma_i$ is the measurement error (see Fig. 1 of Melo 2003, for an example of the procedure).
The CCFs of our objects are plotted in Fig.~1.

For \vsi$>30$ \kms\ the typical error of our \vsi\ measurements 
is 10\%.
For $10<$\vsi$\le30$ \kms\ the error is about 1.5 \kms. 
Our measurements are given in Table~1. 

%------------------------------------------------------------------------------  
 \begin{figure*}  
 \mbox{}  
 \vspace{3.8cm}  
  \includegraphics{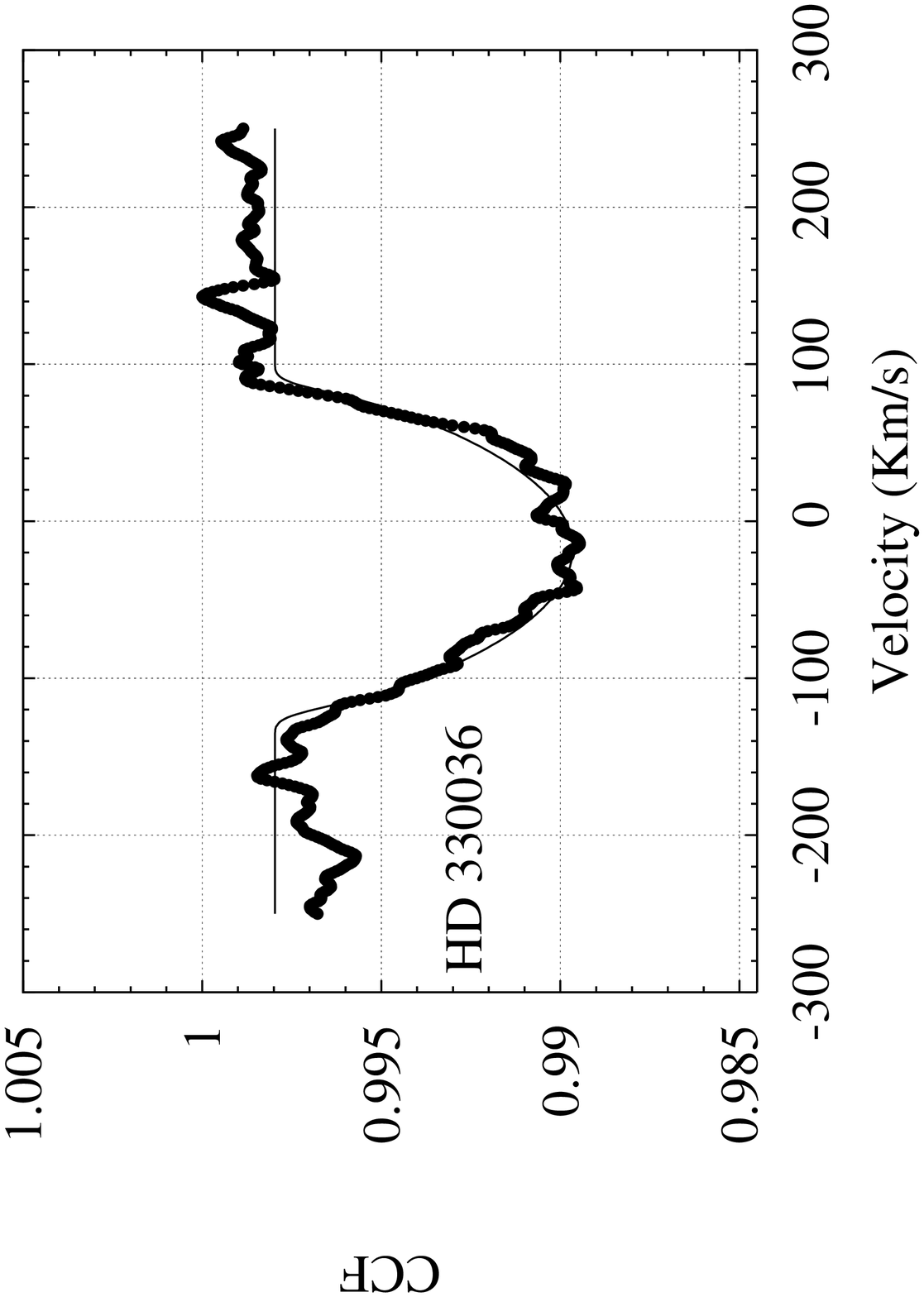}   
  \includegraphics{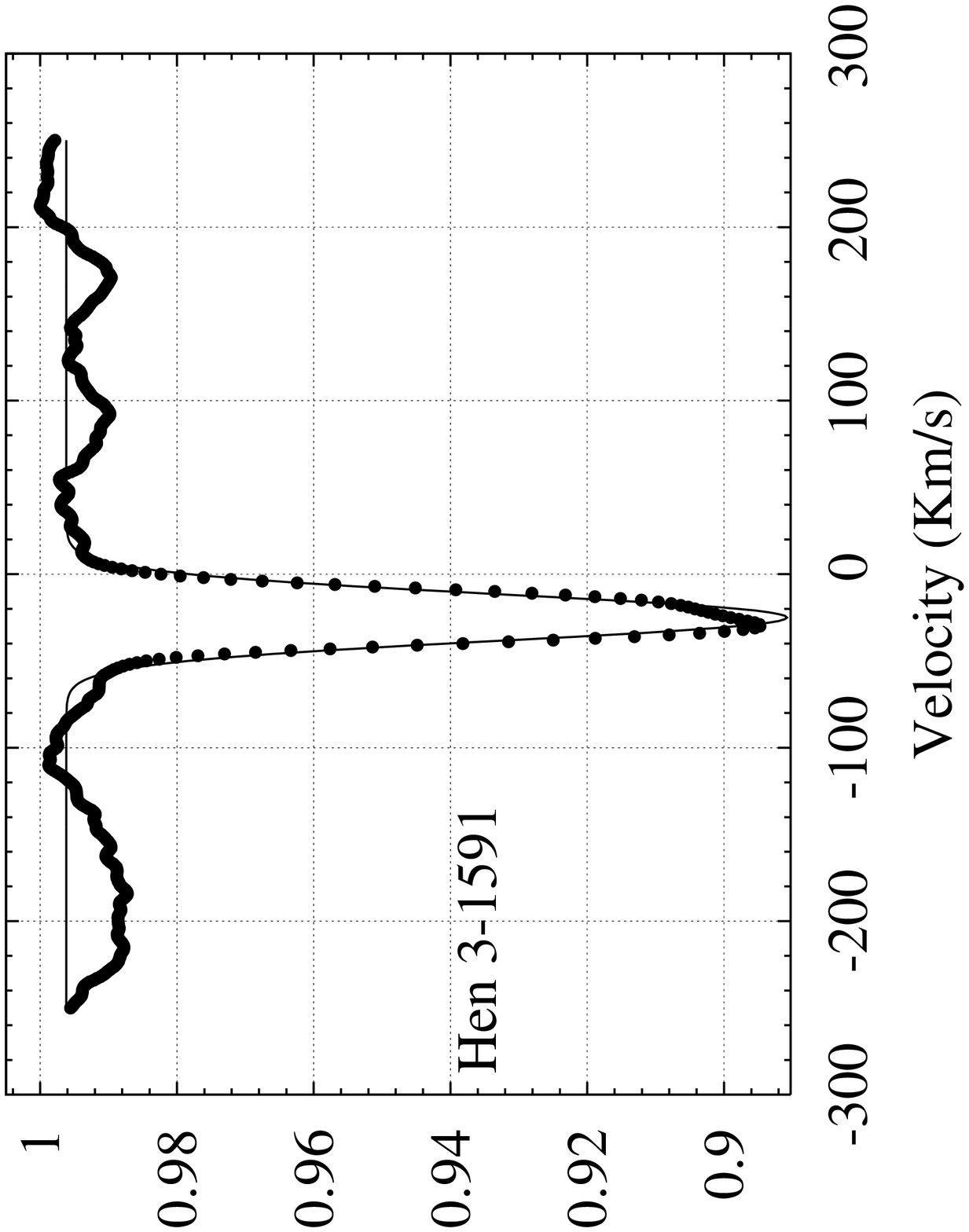}   
  \includegraphics{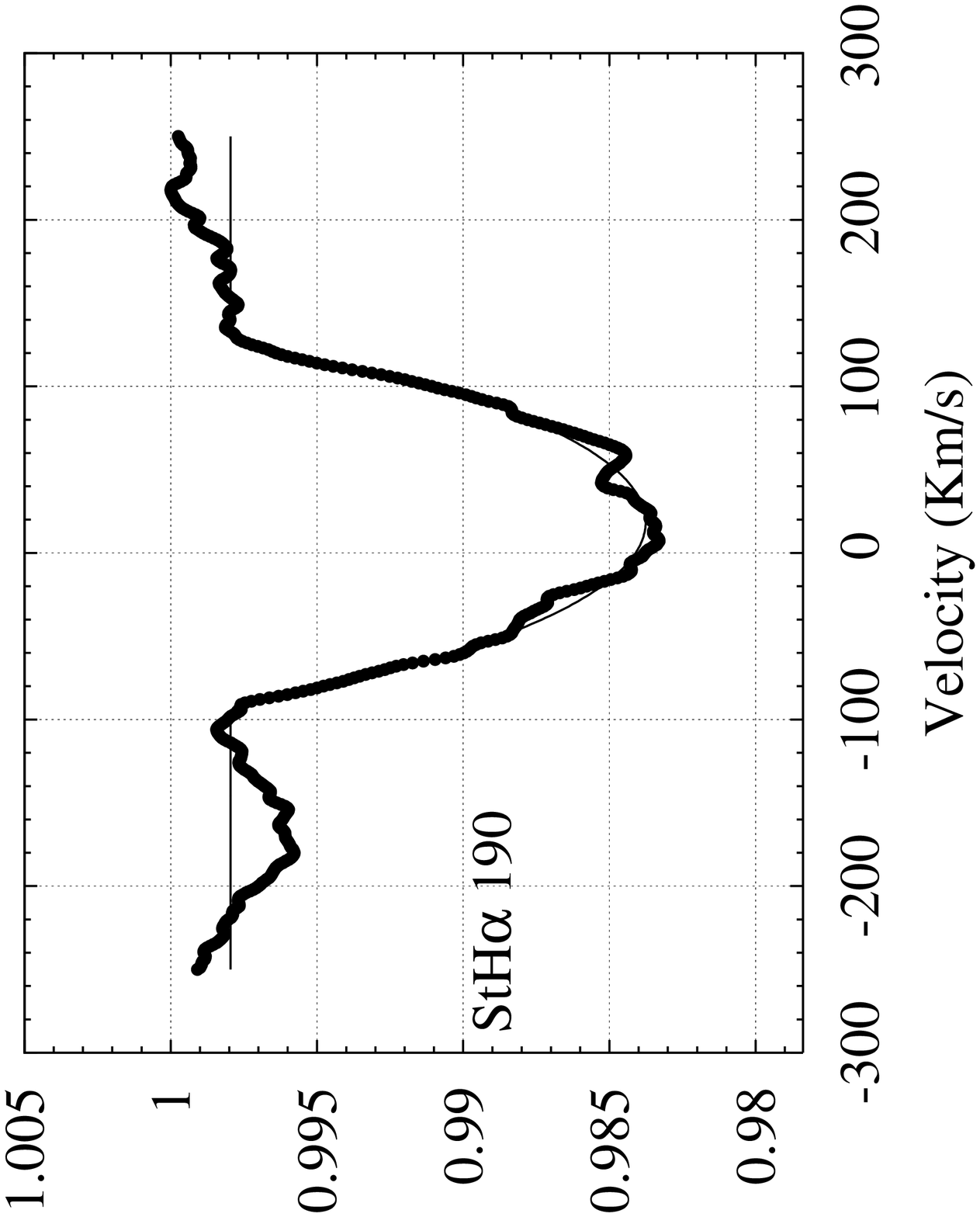}  
  \includegraphics{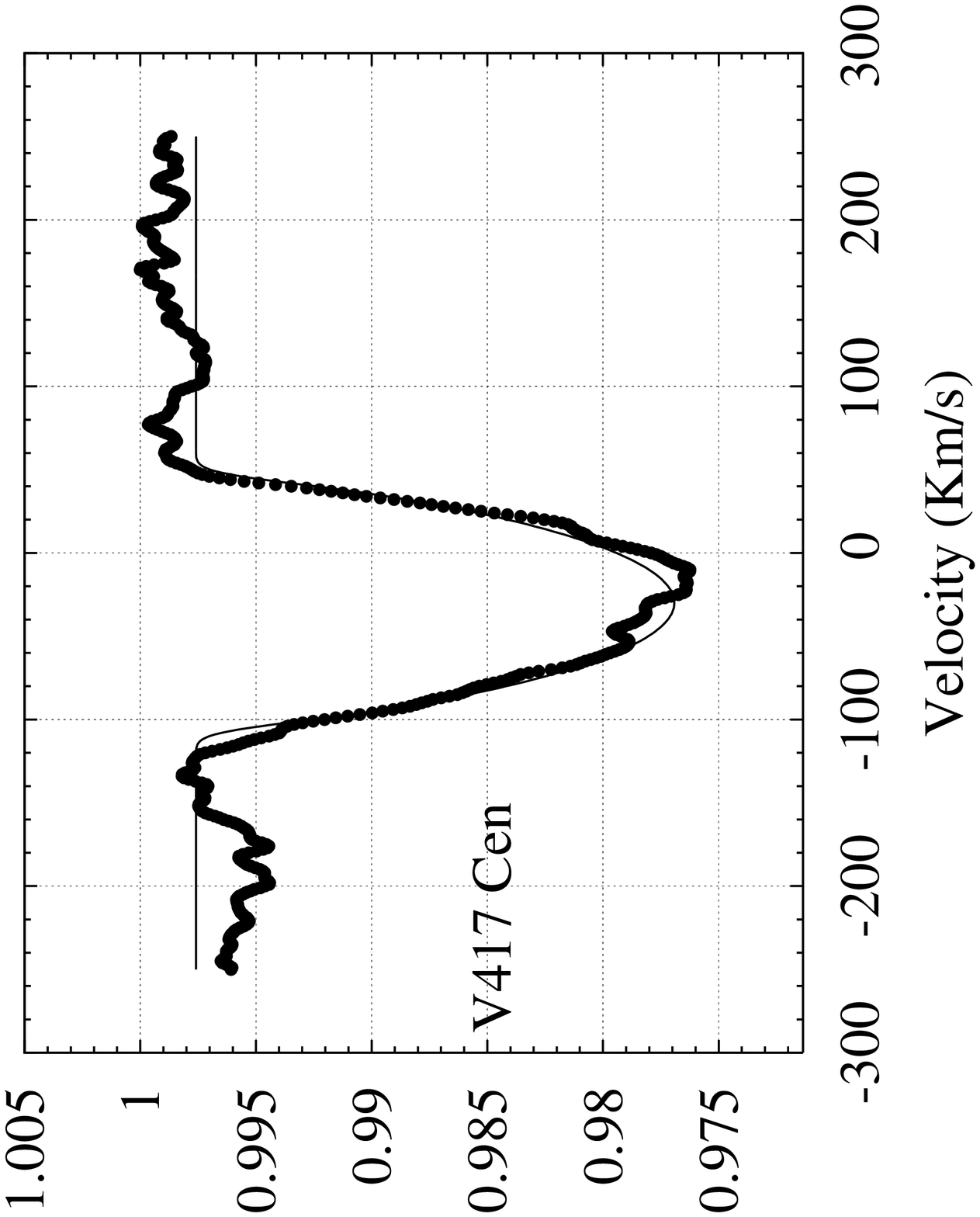}   
\caption[]{ CCF relative intensity (heavy line) and the fit versus
radial velocity for SSs observed in this paper.}	    
\label{CCF}     
\end{figure*}	    
%---------------------------------------------------------------------------------

%  \special{psfile=ccf_xHen-3-1591.eps  hoffset= 60 voffset= 60  hscale=20 vscale=20 angle=270}   
%  \special{psfile=ccf_xStH$\alpha$190.ps      hoffset= 60 voffset=-60  hscale=20 vscale=20 angle=270}   
%  \special{psfile=ccf_xHD330036.ps     hoffset= 60 voffset=-180  hscale=20 vscale=20 angle=270}  
%  \special{psfile= ccf_xV417Cen.ps     hoffset= 60 voffset=-300  hscale=20 vscale=20 angle=270}   
%

\section{Rotation of the mass donors }

%\subsection{Radii of the cool components}
%In Table 2, we also include  R$_g$, the  radius of the cool components.
%For Hen~3-1591, adopting luminosity  class III,
%we will use the average radii for the corresponding  
%spectral type taken  from van  Belle et al.(1999).

%For other three stars, detailed calculations of luminosity,
%and T$_{eff}$ of the cool components 
%are available, we can calculate it directly
%from $L=4\pi \sigma_{SB} R_g^2 T_{eff}^4$. 

\subsection{Individual objects}
\label{indiv}

\begin{table*}
\caption{Parameters of D'--type SS.  
IR types  are from the  catalogue  
of Belczy{\' n}ski et al.(2000), 
\vsi\ is the rotational velocity of the mass donors (as adopted here), R$_g$ and M$_g$ are the radius
and the mass of the giant, respectively (see Sect.\ref{indiv}).  
$\tau_{syn}$ is the synchronization time of the binary
adopting P$_{orb}=$P$_{rot}$,  $\tau_{syn}(100)$ is 
the synchronization time of the binary
adopting P$_{orb}=$100 days, $v_{\rm crit}$
is the  critical rotational velocity of the giant, 
P$_{rot}$ is the rotational period 
of the giant, $a$ is the semimajor axis 
of the system calculated supposing synchronized 
rotation (P$_{orb}=$ P$_{rot}$). The values in brackets for P$_{rot}$  and
$a$ are the upper limits, supposing 10\% uncertainity  in \vsi\ and R$_g$.
}  
  \begin{tabular}{@{}lllrrrrrrccc@{}}
\hline
 object        &IR    &  Cool Star  &\vsi\ & $v_{\rm crit}$& R$_g$      & M$_g$ & $\tau_{syn}$ &$\tau_{syn}(100)$ &P$_{rot}$     & $a$ &        \\
               &type  & Spectrum&[\kms]& \kms   &[R$_{\sun}]$&M$_{\sun}$ & [yr] & [yr] & [days] & [R$_{\sun}$] &      \\
\hline										  &					\\
HD~330036       & D'  &  F8III	&107.0 &160 &	22.1 &  4.46  &25  & 2.1.10$^5$& $<$10.4 (12.8) &      35(40)	  \\  
Hen~3-1591      & S,D'&  K1III	&23.7  &144 &	23.9 &  3.9   &8680& 1.3.10$^5$& $<$51.0 (62.3) &  98(112)    \\   
StH$\alpha$~190 & D'  &G4III/IV &105.0 &191 &	7.88 &  2.25  &27  & 1.3.10$^7$& $<$3.78 (4.64) &  15(17)	 \\   
V417~Cen        & D'  & G9Ib-II &75.0  &105 &	75.0 &  6.45  &36  & 553	  & $<$50.6 (61.8) &  112(128)     \\  
AS~201          & D'  & F9III	&25.0  &150 &	24.5 &  4.35  &6423& 1.1.10$^5$& $<$49.5 (60.5) &      99(113)    \\   
&&&&&& \\								
\end{tabular} 
 \label{D'tab2} 
% \hline       
 \end{table*}

\hskip 0.5cm 
{\bf HD~330036 :}    
Pereira et al. (2005) 
obtained L=650~L$_{\sun}$ for the cool component (with possible uncertainties
160$<L<$3000~L$_{\sun}$), T$_{eff}=$6200\e150~K, $\log g=2.4$\e0.7.
This imply $R_g= 22$R$_{\sun}$ 
(using $L=4\pi \sigma_{SB} R_g^2 T_{eff}^4$),
M$_g=$4.46~$\msun$ (using R$_g$ and $\log\;g$),
 and P$_{rot} \la 10.4$\e2.4~d (using P$_{rot}\;$\vsi$\le 2\pi R_g$).

{\bf  Hen 3-1591 :}  Medina Tanco \& Steiner  (1995)
give spectral type K1 for the cool component. We assume that it is 
luminosity class III. 
The average radius of a K1III star is 23.9\e3~R$_{\sun}$  and 
the average T$_{eff}=$4280\e200~K (van  Belle et al. 1999).
A K1III star would have a mass of 3.9\e0.3 $\msun$ (Allen 1973).
The uncertainties correspond to \e0.5 spectral types.
This will result in L=172~L$_{\sun}$(\e20\%) and 
P$_{rot} \la 51.0$\e11.3~d.

{\bf StH$\alpha$~190 :} The cool component is of type G4 III/IV with T$_{eff}=$5300\e150 K,
$\log g=3.0$\e0.5,
and L=45~L$_{\sun}$ (Smith et al. 2001). This implies R$_g = 7.9\pm0.4$~R$_{\sun}$ 
and  P$_{rot}\la 3.8$\e1.2~days 
(upper limit calculated for R$_g = 8.3$~R$_{\sun}$  and $\sin i=1.0$).
The upper limit for P$_{rot}$ is considerably shorter than 
the supposed orbital periods
of 171~d (Munari et al. 2001) or 38~d (Smith et al. 2001).

{\bf V417 Cen :} Van Winckel et al. (1994) 
detected a photometric period of 245 days.
For  the cool component they obtained G9~Ib-II, 
$\log$ L/L$_{\sun}=$3.5, T$_{eff}=$5000 K, $\log g=1.5$\e0.5. 
This implies $R_g= 75$R$_{\sun}$ and P$_{rot} \la 50.6$\e10.2 days.
P$_{rot}$ is considerably shorter than the period obtained from photometry. 
However, the photometric period 
is not confirmed with radial velocity measurements and we do not know 
whether this is the orbital period. 

{\bf AS 201 :} Following Pereira  et al. (2005), 
the cool component is of type F9III with T$_{eff}=$6000\e100~K,
L=700~L$_{\sun}$ (with possible uncertainties
$300<L<1200$~L$_{\sun}$), and $\log g=2.3$\e0.3. 
This imply R$_g =24.5$~R$_{\sun}$ and 
P$_{rot}\la 49.5$\e11.0~days.  

\subsection{Projected rotational velocities \vsi - comparision with catalogues }

%Soker (2002) has shown  theoretically, that the cool companions in symbiotic systems 
%are likely to rotate much faster than isolated, or in wide binary systems, cool giants. 	   
There are no systematic investigations of the rotation of the mass donors in SSs.
The rotational velocities of 13 S--type systems listed in Fekel et al. (2003)
are between \vsi= 3.6 - 10.4 \kms. All D'--type systems so far observed 
(see Table \ref{D'tab2}) rotate with \vsi$>$20 \kms.	

The catalog of de Medeiros et al. (2002) of \vsi\ of Ib supergiant stars
contains 16 objects from spectral type G8-K0 Ib-II. All they have
\vsi\  in the range 1-20 \kms. It means that V417~Cen
is an extreme case of very fast rotation for this spectral class. 

The  catalogue of rotational  velocities for evolved stars
(de Medeiros et al. 1999) 
lists $\sim$100 K1III stars, and  90\% of them rotate with \vsi$<$8 \kms.
There are only 5 with \vsi$>$20 \kms. This means that
Hen~3-1591 is a very fast rotator (in the top 5\%). 
The same catalog  contains 5 objects from 
spectral type F8III-F9III. They rotate with \vsi\ of 10-35 \kms. 
AS 201 is well within in this range. However HD~330036 is an extremely
fast rotator.
The same catalog lists 60 objects from 
spectral type G3,G4,G5 III-IV. 
They all rotate with \vsi\ $<24$ \kms. 
Again, this means that StH$\alpha$~190 is an extremely fast rotator. 

Thus overall, 4 out of 5 D'--type SSs in our survey are very fast rotators.

%\vskip 0.3cm
\subsection{Critical speed of rotation}
\label{critvel}

There is a natural upper
limit for rotation speeds, where the centripetal acceleration balances that due to
gravitational attraction, often named the ``critical
speed'', where $v_{\rm crit} = \sqrt{GM/1.5R} = 357\sqrt{M/R}$ \kms\
(the factor of 1.5 appears from the assumption that at critical rotation speeds the
equatorial radius is 1.5 times the polar radius, $R$). The calculated $v_{\rm crit}$ is included in Table~\ref{D'tab2}.
No star can rotate faster than its critical speed,
however we can see that at least three D'--type SSs are rotating  at a substantial fraction of
their critical speeds:  
$\frac{v\: \sin i}{v_{\rm crit}}$
$\sim0.67$ (HD~330036),
$\sim0.54$  (StH$\alpha$~190), $\sim0.71$  (V417~Cen). Probably for these three SSs
the orbital inclination is high $i\ge 50^0$.
For the remaining two objects we can not exclude the possibility that
they also rotate very fast but are observed at low 
inclination ($i\la 30^0$).

%\section{Discussion}

\section{ Synchronization  and binary periods}

\subsection{ Synchronization in SS}

The physics of tidal synchronization for stars with convective
envelopes has been analyzed several times (e.g. Zahn 1977 and also the
discussion in Chapter 8 of Tassoul 2000). There are some
differences in the analysis of different authors, leading to varying
synchronization time-scales. Here, we use the estimate from Zahn (1977,
1989): the synchronization time-scale in terms of the period is
\begin{equation}
\tau_{\rm syn} \approx 800 \left( \frac{M_g R_g}{L_g}\right)^{1/3} 
\frac{M_g^2 (\frac{M_g}{M_{WD}} + 1)^2}{R_g^6} P_{orb}^4\ {\rm years}
\end{equation}
where $M_g$ and $M_{WD}$ are the masses of the giant and white dwarf
respectively, and $R_g$ and $L_g$ are the radius and
luminosity of the giant (all in Solar units). The orbital period $P$ is
measured in days.

The S--type SSs are very likely synchronized 
(Schild et al. 2001; Schmutz et al. 1994).
Other proof of this supposition is that most of the SSs with derived orbital parameters
(see  Miko{\l}ajewska 2003) have orbital eccentricity
$e\approx$0. Because the circularization time of the orbits
is $\sim$10 times longer than the synchronization time
(Schmutz et al. 1994 and the references therein), if 
a SS's orbit is circularized it  will very likely be 
synchronized too. 

A typical D'--type SS would have 
$R_g \sim 20\rsun, L_g \sim 500\lsun , M_g \sim 3\msun$, 
and  white dwarf mass $M_{WD} \sim 1\msun$.
For a period of 100 days for a typical D'-type SS, we find 
$\tau_{\rm syn} \sim 9\times 10^4$ years. 

For the individual systems, we calculated the synchronization time ($\tau_{syn}$) 
for two cases:
$\tau_{syn}$ is derived assuming P$_{orb}=$P$_{rot}$,  and  $\tau_{syn}(100)$ 
assuming P$_{orb}=$100 days. These are given in Table~\ref{D'tab2}.
Depending on the individual parameters, the synchronization time
can be from $<$100 yr up to  $>$10$^7$ yr. 
This means that it is possible for a D'--type SS to be synchronized 
if the orbital period is short (P$_{orb}\approx$P$_{rot})$.

% The lifetime of the symbiotic phase for a red giant or AGB star 
% is around $10^5$yr
% (see e.g. Yungleson et al. 1995), and therefore we may expect that
% most of them will also have been tidally synchronized.

\subsection{Evolutionary status of the mass donors}
\label{evolu}

The mass of the mass donors in S--type SSs with known parameters
are in the range 0.6 - 3.2 $\msun$ (Miko{\l}ajewska 2003).
The calculated masses of the mass donors in D'--type SSs
are larger. As can be seen in Table~2, they range from 2.2
up to 6.5 $\msun$.

Masses of 8~M$_{\sun}$ are  generally considered the upper 
limit for evolution to planetary nebula nuclei and white dwarfs, after heavy mass loss,
especially during their AGB %(aasimptotic giant branch)
phases. Following our  calculations 
for the mass of the giants (Table~\ref{D'tab2})  and assuming M$_{WD} \le 1.4\,\msun$, 
the total mass of the binary is about (M$_g+$M$_{WD})\sim 3.5-8.0 \msun$, 
in agreement with the above upper limit of 8~M$_{\sun}$ for the  WD progenitor.

\begin{figure}
 \mbox{}  
 \vspace{9.0cm}  
  \includegraphics{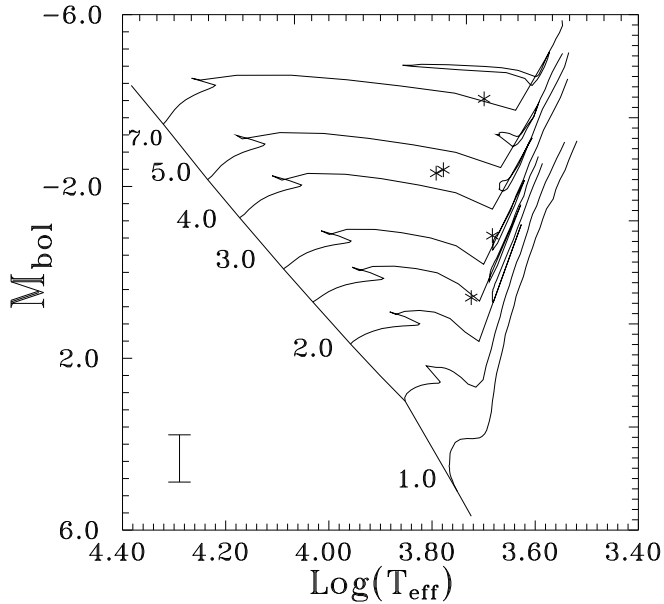}  
  \caption[]{The position of the mass donors on the H-R diagram (see Sect.\ref{evolu}).
  From top to bottom, the objects are V417~Cen, AS~201, HD~330036, Hen~3-1591,  
   StH$\alpha$~190.
  The typical error on the bolometric magnitude, M$_{bol}$, is plotted in the bottom left corner. 
  The typical error on $\log$T$_{eff}$ is \e0.01.   
  }
\label{H-R}     
\end{figure}

The position of the mass donors on the H-R diagram is presented in Fig.\ref{H-R},
assuming near solar chemical composition and  stellar parameters as given in Sect.\ref{indiv}
(see also Pereira 2005).
The evolutionary tracks of Shaller et al. (1992) have been used.
The donors appeared in a wide mass interval -- from 2.5 to $7\msun$. 
The derived evolutionary masses are in good 
agreement with those obtained from R$_g$ and $\log g$.
Three of them appeared
crossing the Hertzsprung gap (HD~330036,V417~Cen and AS~201), 
StH$\alpha$~190 is situated near the base of the red giant branch, and 
He~3-1591 is already evolving on the red giant branch.

The relevant time for a  $5\msun$ star to cross the  Hertzsprung gap
is $\approx$8$\times$10$^5$ yr  and its life time 
on the red giant branch is $\approx$5$\times$10$^5$ yr (Iben 1991).
For a $1.5\msun$  star these times are 1.5$\times$10$^8$ yr  and 1.57$\times$10$^8$ yr 
(Iben 1991). These lifetimes are longer than the calculated $\tau_{\rm syn}$ and comparable with 
$\tau_{\rm syn} (100)$. This means that the rotation of the mass donors in 
D'--type SSs could be synchronized for these lifetimes.  

%%%------------------------------------------------------------------------------  
%-------------------------------------------------------------------------------  

\subsection{Clues regarding the orbital periods}

% The parameters of the D' type SS are summarized in Table~\ref{D'tab2}.
Because the orbital periods of the majority of SSs
are unknown, an indirect method to obtain P$_{orb}$ is to measure 
\vsi.
If the mass donors in SSs are co-rotating (P$_{rot}=$P$_{orb}$), 
we can find clues for the orbital periods via the simple relation 
P$_{orb}$ v$_{rot}$  = 2$\pi$R$_g$, where P$_{orb}$ is the orbital period,
v$_{rot}$ and R$_g$ are the rotational velocity and radius of the giant,
respectively. It is very useful in the case of eclipsing binaries, where 
$\sin i\approx1$ and \vsi$\approx$v$_{rot}$. If the inclination is unknown, we can 
only obtain an upper limit for P$_{orb}$.

Using Kepler's third law 
[$4 \pi a^3 = G (M_g +M_{WD}) P^2  $],
we can calculate the semimajor axis of the systems.
These are given in Table~\ref{D'tab2}. 
The values in
the brackets  (for P$_{rot}$ and $a$) correspond to the estimation of the upper
limit of these parameters, assuming 10\% errors in \vsi\ and R$_g$.

Up to now, from 188 SSs, the orbital elements and binary periods 
are well known for  $\sim$40 objects only (and they are all S--type).
The orbital periods are in the range 200 - 2000 days (Miko{\l}ajewska 2003). 
As can be seen in Table~\ref{D'tab2}, if D'--type SSs
are synchronized, their orbital periods would be 
relatively short (4-60 days) and the distance between
the WD and the mass donor would be 2-5~R$_g$.

\section{Discussion}

%\subsection{Fast rotation}
The observation  of fast rotation of D'--type SSs 
raises two further questions: first, what is the evolutionary
history of these stars which has produced such high rotation, and
second, what does the effect of high rotation have on their mass
loss and subsequent dust formation?
%, subsequent evolution and dust formation.

We see three possible reasons for the fast rotation of mass donors in D'--type SSs:

{\bf (i)} the rotation is synchronized with the binary period 
(P$_{rot}=$ P$_{orb}$).
In this case their orbital periods  should  be short $\la$50 days. 

However, it also could be, that  they  are not synchronously  
rotating. The possibility that 
P$_{rot}>$ P$_{orb}$ has to be excluded because the orbital
separations would be unreasonably small, and the synchronization time would be 
extremely short (they will be synchronized in 30-9000 years; 
see $\tau_{syn}$ in Table~\ref{D'tab2}).
If P$_{rot}<$ P$_{orb}$,
reasons for their fast rotation could be:

{\bf (ii)} the  current giants 
have been spun-up from the transfer of angular momentum.
Jeffries \& Stevens (1996) proposed a mechanism in which 
accretion of a slow massive wind from the AGB progenitor of the 
current white dwarf can transfer sufficient angular momentum. 
This also explains the chemical enrichment in s-process elements
in D' SSs, 
that were present in the AGB wind (see also Pereira et al. 2005). 
Mass transfer via L$_1$, 
when the current white dwarf was the mass donor, 
can also spin-up the companion, as in millisecond radio pulsars
(van den Heuvel 1984).

{\bf (iii)} planet swallowing to spin up the giant. 
Rough estimation gives angular momentum transfer during a collision of a planet with mass $m_p$ 
and velocity $v_p$ to a giant of mass $M_g$, of the order of
$\Delta \Gamma = m_p v_p R_g$. This causes a change in the giant's rotational velocity
$\Delta v^{rot}_g = {m_p v_p}/{const. M_g}$, where const.=$J_g/M_g R_g^2$ depends on the internal structure of the giant.
($const.=0.4$ for a solid uniform density sphere, and less for a star-like centrally condensed sphere). 
Assuming a centrally condensed star (giant) such that $const.=0.01$, $v_p = \sqrt{GM_g/R} \sim 10 \-- 100$ \kms, 
$M_g=2\--10 M_\odot$ 
and $m_p=0.01 M_\odot$ we estimate $\Delta v^{rot}_g \sim 1 \-- 50$ \kms, showing that in the right circumstances, the planet 
could spin up the giant to the rotational velocities observed in D'--type SSs.

\vskip 0.3cm
%\subsection{Mass loss}
Fast rotation, i.e. $v_{rot}\ge0.5{v_{\rm crit}}$,  may change 
a spherical star with a spherical wind into an equatorially flattened system,
with both the radius of the star and stellar wind parameters depending on the stellar latitude.
Such stars will have an equatorial radius significantly larger than the polar one, and 
equatorially enhanced mass loss (see Lamers 2004 and references therein).

Because it seems that the majority of the giants in 
D'--type SSs are rapid rotators (Sect. \ref{critvel}), 
we  expect that: \\
{\bf (1)} they have a larger mass loss rate than the slower rotating giants;\\
{\bf (2)} their mass loss is enhanced in the equatorial regions;\\
{\bf (3)} they could be even  equatorially flattened.  \\

It is possible, that the dusty environment in D'--type SSs is connected with 
rapid rotation of the mass donors. 
Intense mass loss in the equatorial regions can 
lead to the formation of an excretion disc in which the higher gas density enhances dust formation and growth. 
%The outflowing wind of gas 
%molecules expands and cools and dust grains can condense in 
%such an outflowing disc.
The broad IR 
excess in D'--type SSs can be due to the temperature stratification 
in the dust from such an excretion disc (Van Winckel et al. 1994). 
% In this way, the intense mass outflow in equatorial regions may lead to 
% dust formation.
Other possible explanations for the presence of  dust can be that
it 
is left over from the formation of a planetary system, or  it is a relic from 
a strong dusty mass loss when the present day white dwarf 
was on the AGB. 
However, we consider it is more likely that
it originates in the current outflow and that this is enhanced in the
equatorial regions by rapid rotation.

\section{Conclusions}  
 
Our main results are:

(1) We have measured the rotational velocities of the mass donors in D'--type
symbiotic stars, using the CCF approach. 

(2) Four of the five objects  appeared to be very fast rotators compared 
with the catalogues of \vsi\ for the corresponding spectral types.
At least three of them rotate at a substantial fraction ($\ga0.5$) of
the critical velocity.
This means that at least in D'--type SSs, the cool components rotate
faster than isolated giants (as predicted by Soker 2002).

(3) If they are tidally locked, the orbital periods 
should be as short  as $\la$10-50 days.

(4) As a result of the rapid rotation, they must have
larger mass loss rates than the more slowly rotating giants, 
and their mass loss is probably enhanced in the equatorial regions.

% This 
% could create appropriate conditions for formation of dust 
% in the outer regions of the equatorial wind.

To understand better these objects, we need their binary periods
to be derived from radial velocity measurements and the inclination determined.
In subsequent papers, we plan to explore the projected rotational
velocity of the cool giants in the other types of symbiotic stars and 
to compare their rotational velocities with that of the isolated giants 
with similar mass and evolutionary stage. 

\section*{Acknowledgments}  
  
This research has made use of MIDAS, IRAF, SIMBAD and  Starlink.  
RZ is supported by a PPARC Research Assistantship  
and MFB is a PPARC Senior Fellow. 
We also acknowledge the vital contribution made by Dr John Porter 
to our successful telescope time proposals and to this paper, 
which was underway at the time of his tragic death.


\begin{thebibliography}{99} 

\bibitem[Allen(1973)]{1973asqu.book.....A} Allen, C.~W.\ 1973, London: 
University of London, Athlone Press, |c1973, 3rd ed.,  
 
\bibitem[Belczy{\' n}ski et ll.(2000)]{2000A&AS..146..407B}   
Belczy{\'n}ski, K., Miko{\l}ajewska, J., Munari, U., Ivison, R.~J., 
\& Friedjung, M.,  2000,  A\&AS, 146, 407   

\bibitem[Corradi et al.(2003)]{2003ASPC..303.....C} Corradi, R.~L.~M., 
Mikolajewska, J., \& Mahoney, T.~J.\ 2003, Symbiotic Stars Probing Stellar Evolution,
Astronomical Society of the  Pacific Conference Series, 303  

\bibitem[de Medeiros \& Mayor(1999)]{1999A&AS..139..433D} de Medeiros, 
J.~R., \& Mayor, M.\ 1999, A\&AS, 139, 433

\bibitem[De Medeiros et al.(2002)]{2002A&A...395...97D} De Medeiros, J.~R., 
Udry, S., Burki, G., \& Mayor, M.\ 2002, A\&A, 395, 97 


\bibitem[Delfosse et al.(1998)]{1998A&A...331..581D} Delfosse, X., 
Forveille, T., Perrier, C., \& Mayor, M.\ 1998, A\&A, 331, 581 

\bibitem[Fekel et al.(2003)]{2003ASPC..303..113F} Fekel, F.~C., Hinkle, 
K.~H., \& Joyce, R.~R.\ 2003, Astronomical Society of the Pacific 
Conference Series, 303, 113 
 

\bibitem[Gray(1976)]{1976oasp.book.....G} Gray, D.~F.\ 1976, 
The observation and analysis of stellar photospheres, Research 
supported by the National Research Council of Canada.~New York, 
Wiley-Interscience, 1976.~484 p.

\bibitem[Iben(1991)]{1991ApJS...76...55I} Iben, I.~J.\ 1991, ApJS, 76, 55 

\bibitem[Kaufer et al.(1999)]{1999Msngr..95....8K} Kaufer, A., Stahl, O., 
Tubbesing, S., Norregaard, P., Avila, G., Francois, P., Pasquini, L., \& 
Pizzella, A.\ 1999, The Messenger, 95, 8 

\bibitem[Kenyon(1986)]{1986syst.book.....K} 
Kenyon, S.~J.\ 1986, The symbiotic stars, 
Cambridge and New York, Cambridge University Press, 1986, 295 p.,  

\bibitem[Lamers(2004)]{2004IAUS..215..479L} Lamers, H.~J.~G.~L.~M.\ 2004, 
IAU Symposium, 215, 479

\bibitem[Medina Tanco \& Steiner(1995)]{1995AJ....109.1770M} Medina Tanco, 
G.~A., \& Steiner, J.~E.\ 1995, AJ, 109, 1770 

\bibitem[Melo et al.(2001)]{2001A&A...375..851M} Melo, C.~H.~F., Pasquini, 
L., \& De Medeiros, J.~R.\ 2001, A\&A, 375, 851 

\bibitem[Melo(2003)]{2003A&A...410..269M} Melo, C.~H.~F.\ 2003, A\&A, 410, 
269 

\bibitem[Miko{\l}ajewska(2003)]{2003ASPC..303....9M} Miko{\l}ajewska, J.\ 
2003, Astronomical Society of the Pacific Conference Series, 303, 9 

\bibitem[Munari et al.(2001)]{2001A&A...369L...1M} Munari, U., et al.\ 
2001, A\&A, 369, L1 

\bibitem[Pereira et al.(2005)]{2005A&A...429..993P} Pereira, C.~B., Smith, 
V.~V., \& Cunha, K.\ 2005, A\&A, 429, 993 

\bibitem[Smith et al.(2001)]{2001ApJ...556L..55S}
Smith, V.~V., Pereira, C.~B., \& Cunha, K.\ 2001, ApJ, 556, L55 

\bibitem[Tassoul(2000)]{2000stro.book.....T} Tassoul, J.\ 2000, Stellar 
rotation / Jean-Louis Tassoul.~Cambridge ; New York : Cambridge University 
Press, 2000.~(Cambridge astrophysics series ; 36),  


\bibitem[van Belle et al.(1999)]{1999AJ....117..521V} van Belle, G.~T., Lane, B.F., 
Thompson, R.R., Doden, A.F., Colavita, M.M.,  et  al.\ 1999, AJ, 117, 521 

\bibitem[van den Heuvel(1984)]{1984JApA....5..209V} van den Heuvel, 
E.~P.~J.\ 1984, Journal of Astrophysics and Astronomy, 5, 209 
 
\bibitem[Van Winckel et al.(1994)]{1994A&A...285..241V} Van Winckel, H., 
Schwarz, H.~E., Duerbeck, H.~W., \& Fuhrmann, B.\ 1994, A\&A, 285, 241 

\bibitem[Schaller et al.(1992)]{1992A&AS...96..269S} Schaller, G., 
Schaerer, D., Meynet, G., \& Maeder, A.\ 1992, A\&AS, 96, 269 
 
\bibitem[Zahn(1977)]{1977A&A....57..383Z} Zahn, J.-P.\ 1977, A\&A, 57, 383 

\bibitem[Zahn(1989)]{1989A&A...220..112Z} Zahn, J.-P.\ 1989, A\&A, 220, 112 

%\bibitem[Zombeck(1990)]{1990hsaa.book.....Z} Zombeck, M.~V.\ 1990, 
%Cambridge: University Press, 1990, 2nd ed.,  






\end{thebibliography}
\end{document}